\documentclass[twocolumn,nofootinbib]{revtex4}
\usepackage{graphicx}
\begin{document}

\title{Comment on ``Are the spectra of geometrical operators in Loop \\ Quantum
Gravity really discrete?" by B.\,Dittrich and T.\,Thiemann}

\author{Carlo Rovelli}
\affiliation{Centre de Physique Th\'eorique de Luminy, Case 907, F-13288 Marseille, EU}
\date{\today}
\begin{abstract}
\noindent 
I argue that the prediction of 
physical discreteness at the Planck scale in loop gravity 
is a reasonable conclusion that derives from a sensible ensemble 
of hypotheses, in spite of some
contrary arguments considered in an interesting recent
paper by Dittrich and Thiemann.  The counter-example 
presented by Dittrich and Thiemann 
illustrates a pathology which does not seem to be present in 
gravity.  I also point out a common confusion between
two distinct frameworks for the interpretation of 
general-covariant quantum theory, and observe that within one 
of these, the derivation of physical discreteness 
is immediate, and not  in contradiction with gauge invariance.

\end{abstract}

\maketitle

\section{Introduction}

In loop quantum gravity (LQG) \cite{libroT, libro, lqg1, lqg2}, 
some operators representing geometrical  quantities have 
discrete spectra \cite{lqg}.   Does it follow that LQG predicts 
that a  Plank-scale measurement of these quantities  would
yield quantized outcomes? That is, does LQG truly predicts
discreteness of physical space? 
These operators are not themselves gauge-invariant;  
but it has been argued that under plausible hypotheses their 
spectra provide evidence for physical discreteness \cite{disc}. 
The plausibility of these additional hypotheses has raised a 
certain debate in the past. 

In a recent paper \cite{DT} Dittrich and Thiemann reconsider the 
debate and argue for a negative answer. They present some
toy models as counter-examples to the expectation that
a discrete  spectrum implies physical quantization.

Here, I argue that the existence of these counter-examples makes 
an interesting point, but  does not change the fact that physical 
discreteness in LQG is a reasonable conclusion deriving from a 
sensible ensemble of hypotheses. There are three distinct arguments
to this effect

(i) As discussed below, the key counter-example presented in \cite{DT} 
is based on a pathology which does not appear to be present in 
gravity, at least in this form.  

(ii) Physical evidence is distinct from   
mathematical proof.   Physics requires     
tentative generalizations and implicit assumptions. 
In loop gravity, evidence is towards 
discreteness, and in support of the familiar relation 
between discrete spectra and quantization.  Lack 
of a rigorous mathematical proof does not imply 
lack of evidence.\footnote{%
In this regard, a stronger wording in the first online 
version of \cite{DT}, (``fundamental discreteness in LQG 
is an empty statement") has been removed in the second 
version.}

(iii) Most importantly, the analysis in  \cite{DT} is based 
on a certain specific framework for interpreting
general-covariant quantum theory.  Dittrich and Thiemann 
refer to my own work on the subject, in particular to the notion of  
``partial observable" and the paper  \cite{partial}.   However, 
ideas discussed in  \cite{partial} are related to an interpretation
which is different from the one they use. Under this second
interpretation physical discreteness follows immediately, 
without contradicting gauge invariance.  

The reason why there exist different  interpretations of 
general-covariant quantum theory, is that we do not know the 
final quantum theory of gravity yet.  We not only lack a confirmed 
mathematical framework, but also, perhaps even more 
importantly, a confirmed physical interpretation for general-covariant 
quantum theory.   In the standard interpretation of quantum theory, a 
quantity is predicted to have discrete values if the corresponding 
\emph{kinematical} operator has discrete 
spectrum. 
How does this translate to a general-covariant theory, where kinematics 
and dynamics are less clearly separated?  There is more than one 
possible answer to this question, and there are 
different ways of  extending the well-tested 
interpretation of standard quantum mechanics to the unknown 
quantum-gravitational regime.  If we interpret 
LQG as suggested in \cite{partial} and in \cite{libro}, then physical 
discreteness is immediate. 

The interest of the Dittrich-Thiemann paper is 
that it shows how distinct ways of extending the interpretation
might yield different physical predictions.  This is an important contribution, 
which deserves to be better understood. 
But in either case, the evidence remains strong towards the 
conclusion that  LQG implies fundamental discreteness at Planck scale.

I point out the difference between two 
frameworks for interpreting general-covariant quantum theory
in Sec.\,II.  The  counter-example in \cite{DT} is discussed in 
Sec.\,III. Some general 
arguments about discreteness of space in LQG are
recalled in Sec.\,IV.  In the conclusion, I summarize the evidence for 
discreteness in LQG.

\section{Two interpretations of general-covariant quantum theory}

There is a common confusion between two distinct  
approaches to the issue of observability in general-covariant quantum
systems.  (In my own work I have studied both of them, beginning with the first, 
and then shifting to the second.)
\begin{itemize}
\item[I.]  
The formulation based on the notion of  ``evolving constant" or
``complete observables" 
discussed in \cite{ec1,ec2,ec3}. This is the interpretation 
that Dittrich and Thiemann utilize. 
\item[II.]    The formulation derived from  the 
paper \cite{partial} on partial observables, and 
summarized for instance in \cite{libro}.  
\end{itemize}
Let me briefly summarize the two, in order to clarify their difference. 
I give here only the basic structure of the 
two interpretations. Both have been discussed in depth in the literature. See 
the references cited for motivation and for the numerous details needed to 
make these pictures physically and mathematically precise.  I assume here 
for simplicity that we are dealing with systems defined by a single 
constraint. 

\begin{itemize}
\item[I.]  
 A general-covariant system is defined by a phase space $\Gamma$, which
is equipped with Poisson brackets. Its dynamics is determined by a (``hamiltonian")
constraint $H$, which is a real function on $\Gamma$.   The quantities that
represent physical
measurements are real functions $F$ on $\Gamma$ that have vanishing Poisson
brackets with $H$ on the surface $H=0$. These are called Dirac observables. 
A special class of these functions is given by the ``evolving constants". An
evolving constant $F_\tau$ is a one-parameter family of Dirac observables 
determined by two functions on $\Gamma$: a ``clock" function $t$ and a
function $f$.  $F_\tau$ is defined by two properties: (i) it is a Dirac observable
and (ii) 
\begin{equation}
(F_\tau=f)|_{t=\tau},
\end{equation}
that is, $F_\tau=f$ on the surface $t=\tau$ in $\Gamma$.  The physical interpretation
of the evolving constant $F_\tau$  is that it represents ``the value of $f$ when the `clock' variable
$t$ has the value $\tau$".  However, notice that in this framework one assumes that
there is no meaning to measuring $t$ by itself, or $f$ by itself.\\
The quantum theory is defined by a kinematical Hilbert space $\cal K$, which
quantizes $\Gamma$, and a constraint operator $\hat H$. The space of the (possibly 
generalized) states that
satisfy $\hat H\psi=0$ is the physical state space $\cal H$ and inherits a scalar product
from $\cal K$. Dirac observables, and in particular evolving constants,  
become operators on $\cal H$.  Transition amplitudes are 
given by the scalar product in $\cal H$ between their eigenstates.  
Physical discreteness is determined by the spectra of the Dirac observables.
In particular, the ``measurement of  the value of $f$ when the 
`clock' variable $t$ has the value $\tau$"  is predicted 
to have quantized outcomes if and only if the operator $\hat F_\tau$ has 
discrete spectrum. 

\item[II.]  
 A general-covariant system is defined by a set of kinematical 
quantities $q_n$ called ``partial observables".  These are physical 
quantities that can  be measured, but are not necessarily 
predictable \cite{partial}. (An example of a quantity considered ``measurable 
but not predictable" is the usual time variable $t$). 
The space of the partial observables is called the extended 
configuration space $\cal C$.  Dynamics 
is given by a (hamiltonian) constraint $H$ 
on $\Gamma=T^*{\cal C}$. \\
The quantum theory is defined by the kinematical Hilbert space $\cal K$ 
that quantizes $\Gamma$ and by the (possibly generalized) ``projection" 
operator 
\begin{equation}
P: {\cal K} \to {\cal H}
\end{equation}
to the space $\cal H$ of the solutions of $\hat H\psi=0$.  The probability amplitude 
for observing the values $q_n$ if the values $q'_n$ 
have been observed is given by 
\begin{equation}
\langle q_n|P|q'_n \rangle,
\end{equation}
where 
$|q_n \rangle$ is the eigenvector of the partial-observable
operator $\hat q_n$ in $\cal K$. If the operator $\hat q_n$ on $\cal K$ has 
discrete spectrum, then the measurement of  the corresponding 
partial observable is predicted to have quantized outcomes.
\end{itemize}

Both frameworks reduce to the standard interpretation of 
quantum mechanics when applied to a conventional 
quantum  system.  In this case $q_n=(q_0,q_a)=(t,q_a)$ where
$q_a$ are coordinates of the usual configuration space, and 
$H=p_0+H_0(p_a,q_a)$, where $p_n$ are the momenta conjugate 
to $q_n$ and $H_0(p_a,q_a)$ is the conventional 
hamiltonian. The quantum operators $\hat q_a$ and $\hat p_a$ are the
usual Schr\"odinger operators, while the evolving-constant 
operator $(\hat Q_a)_\tau$ determined by $q_a$ and $t$ is 
the Heisenberg operator 
\begin{equation}
(\hat Q_a)_\tau=e^{-i\hat H_0\tau }\hat q_ae^ {i\hat H_0\tau }. 
\end{equation}
Notice that the operators $(\hat Q_a)_\tau$ and $\hat q_a$ have the same
spectrum, because $e^{-i\hat H_0\tau }$ is unitary.  Hence $(\hat Q_a)_\tau$
has discrete spectrum if and only if  $\hat q_a$ does. 

In the general case, the two interpretations are often compatible,
but not always. The paper \cite{DT} interestingly points out a number of cases 
where the two appear to lead to different predictions concerning 
discreteness.   Which one is then the correct generalization of 
quantum theory?   I shall not enter this discussion here (a few comments are below, in Sec.\,III). I only mention that my own opinion has somehow shifted over the years from I to II, and  I  
have given arguments for this in \cite{libro}.  

What is relevant for the present 
discussion is the following.  The geometrical operators that have discrete spectra 
in LQG can be considered partial observables. If one follows interpretation II,  
the conclusion that a physical measurement of these quantities yields 
quantized values is immediate, because physical quantization depends on
the spectra of \emph{kinematical} operators in $\cal K$ in this interpretation. 

If, instead, one follows interpretation I, as  Dittrich and Thiemann, then the
possibility is open in principle that the spectrum of a quantity $f$ be 
discrete but a corresponding complete observable $F_\tau$ for some
clock time $t$ has continuous spectrum.  

For instance, it is in principle possible that the partial observables $\hat f$ giving
the area of a coordinate surface has discrete spectrum,
while the corresponding complete observable $\hat F_\tau$ representing the area
of a physically defined surface, has continuous spectrum. 
This is the possibility that Dittrich and Thiemann point out. It is
a possibility. Is it a plausible one, for the geometrical quantities in LQG? 

\section{The Dittrich-Thiemann counter-example}

In \cite{DT}, various examples are presented where the quantity $f$ is
represented by an operator $\hat f$ in $\cal K$ which has discrete
spectrum, while a corresponding evolving constant operator 
$\hat F_\tau$ does not.  As stated in \cite{DT},
most of these examples can be taken as irrelevant for the situation in LQG 
because they differ substantially from the LQG case.
First, the configuration space of the models is non compact,
while for the geometry part of LQG the configuration space is compact. 
Second, $f$ and $t$ do not commute, which is a different situation
than the one expected in LQG.  But in  \cite{DT} there is also an 
example presented as a ``baby version" of LQG, meant to
show what could effectively go wrong  in LQG in expecting 
a discrete spectrum of
a kinematical operator to naturally imply a discrete spectrum of a 
corresponding complete observable.

The model is defined by the phase space $T^*(S^1\times R)$,
namely the cotangent bundle of a cylinder.  
The angular variable of the
cylinder can be called\footnote{I change notation with respect to \cite{DT}, 
because I find confusing using a Greek letter for a variable in $R$ and a 
Latin letter for an angular variable.} 
$\alpha$, by analogy with the gravitational connection $A$.
Since it varies on the compact space $S^1$, its conjugate variable $p_\alpha$
is given in the kinematical state space $\cal K$ by an operator with
discrete spectrum. The longitudinal variable of the cylinder can be called 
$x$, and its conjugate momentum $p_x$.   Hence: $\alpha\!\in\! S_1$, $x\!\in\! R$.  
Dittrich and Thiemann choose a dynamics  defined by a hamiltonian constraint 
that can be written in the form
\begin{equation}
                        H = \cos(\alpha-x)-1=0 
                        \label{H}
\end{equation}
and show that the spectrum of the evolving constant $(P_\alpha)_\tau$ determined
by $p_\alpha$ and by the clock variable $t:= p_x$ is continuous. More precisely, we
can avoid the ill-defined angular variable $\alpha$ by defining 
\begin{equation}
                        h = e^{i\alpha}  
                        \label{h}
\end{equation}
in $S_1$, and the constraint reads then 
\begin{equation}
                        H= h - e^{ix}  =0.  
                        \label{H2}
\end{equation}
(this should be multiplied by $i$ to generate a real evolution.)
As Dittrich and Thiemann 
nicely explain, (\ref{H2}) constrains the cylinder down to a spiral in it (see Fig.\,1), and the 
dynamical disappearance of the quantization of $p_\alpha$ is due to the following fact.
While its conjugate variable $\alpha$ varies over a \emph{compact} space $S_1$
in the unconstrained phase space, the variation of $\alpha$ gives a flow along the 
\emph{non-compact} spiral, when we follow it along the constraint surface. 

More precisely. Recall that  a characteristic indication that an operator will 
have discrete spectrum is that 
the hamiltonian flow generated by the corresponding classical quantity is compact. The flow
generated by $p_\alpha$ over $T^ *(S^1\times R)$ is compact: it is an angular rotation of the cylinder. But there can be no non-trivial compact flow over the spiral.  

\begin{figure}
\begin{center}
  \includegraphics[height=4cm]{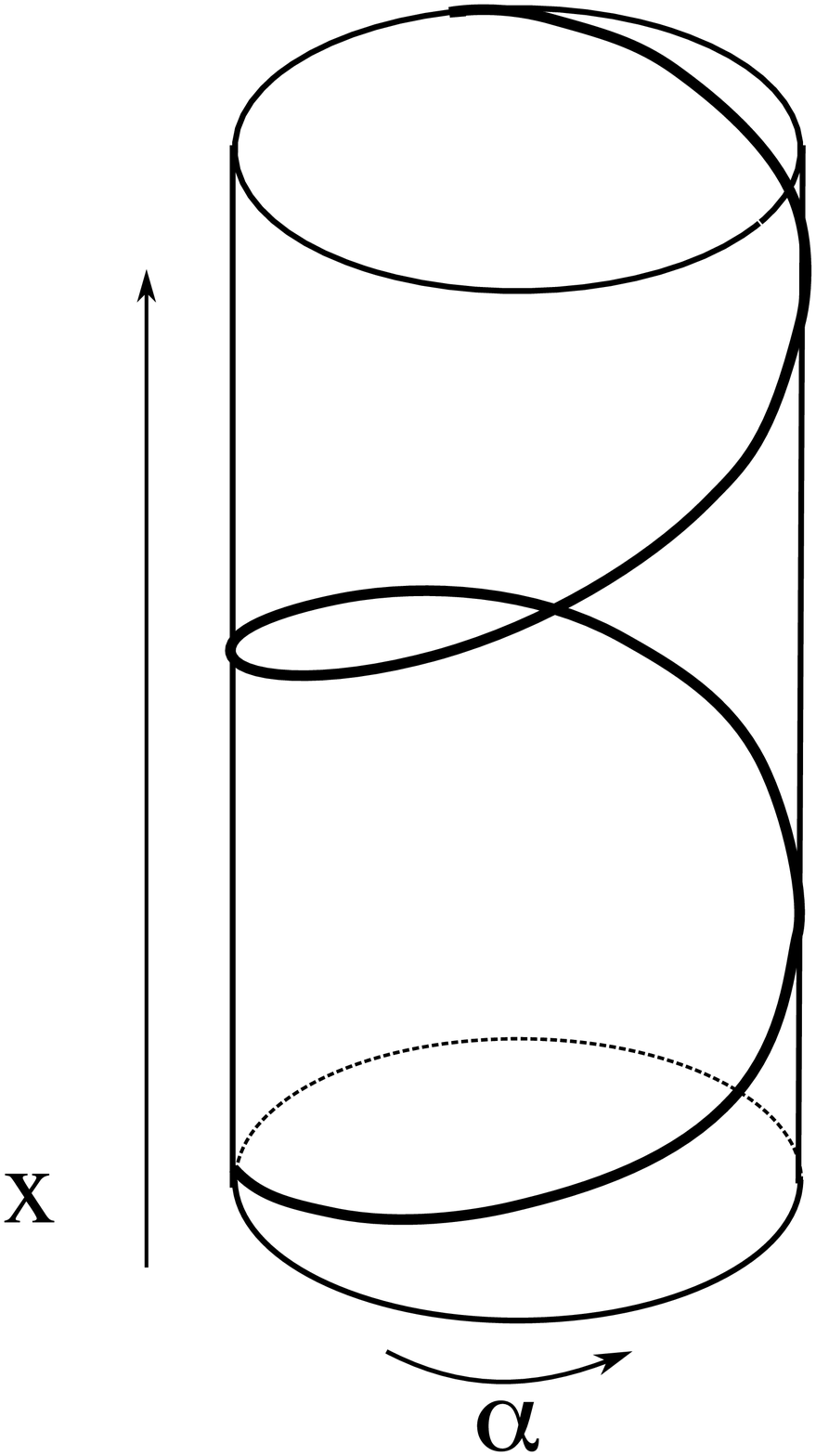}
\end{center}\vspace{-2em}
\caption{The configuration space (cylinder) and the constrained configuration space (spiral).}
\label{spiral}
\end{figure}

Notice that this happens because the constraint (\ref{H2}) is a 
\emph{periodic} function of $x$. In this way, the same value of $h$ 
determines an infinite set of disconnected values of $x$, and therefore 
(\ref{H2}) determines a one-to-many relation from $S_1$ to the spiral. The
compact flow along $S_1$ is then \emph{necessarily} ``opened-up" to
a non-compact flow along the spiral.
In particular, there is no embedding of $S_1$ into the constraint
surface  (\ref{H2}). Any circle $S^1$ is necessarily ``opened-up" by the 
constraint  (\ref{H2}).  This is what destroys the quantum discreteness.

As far as quantum gravity is concerned  the relevant question is then whether the 
same can happen in general relativity. Could there be a similar mechanism that
forbids the compact action of a flow? The gravitational variables used 
in LQG are holonomies of the gravitational $SU(2)$ Ashtekar-Barbero 
connection $A$. Consider one such holonomy, and call it $h$.
It lives in the compact space $SU(2)$.  The gravitational constraints
can be expressed in terms of the holonomies \cite{libroT}.  Can it be
that varying $h$ in $SU(2)$ along the constraint surface gives necessarily 
a non-compact flow, as in the ``baby version" above?  For this to happen, there 
should be no embedding of $SU(2)$ in the constraint surface of general
relativity, possibly coupled with matter.  

But say we consider GR plus a 
set of scalar fields.  The constraints are then given by the standard
gravitational constraints plus the matter energy, or momentum. Thus,
we can always solve the constraints by simply choosing an initial
configuration of matter with given energy and momentum. But the
dependence of these on the matter degrees of freedom is linear or
quadratic, and not a periodic, or more complicated function.  Hence 
there is no obvious way of forcing the $SU(2)$ flow obtained  
by varying $h$ to  open-up to  a non-compact one on the constraint 
surface, as in the counter-example.   It is thus reasonable
to expect discreteness to survive for some clock choice.

Of course, this does not prove that complete observables corresponding to
physically determined areas have discrete spectrum in LQG. But it shows 
that the counter-example presented depends on a peculiar pathology that
does not appear to be reproduced in general relativity.

\section{Arguments for discreteness}

I review some general physical arguments in favor of the conclusion
that LQG predicts physical discreteness, beginning with those in favor 
of interpretation II. \\ 

(i) {\em Kinematical character of quantum discreteness.} In conventional quantum theory, the discreteness of a physical variable
is a {kinematical} property. It is independent
from the dynamics. 

For instance, the momentum $p_\varphi$ of
a particle that moves on a circle, namely whose position is given by an
angular variable $\varphi\in S_1$, is quantized \emph{whatever} is the hamiltonian
$H_0(\varphi,p_\varphi)$. I find it hard to believe that this {kinematical} character of quantum 
discreteness does not survive in a general-covariant theory.

 The 
distinction between kinematics and dynamics becomes subtle
in a general-relativistic theory, but it still exists.   For instance, 
consider the general-relativistic dynamics of the 
solar system. Kinematics determines the relevant variables 
(distances of the planets, angles in the sky, various proper times...); while
the expected relations that the equations of motion of the theory impose 
among these variables 
constitute the dynamics.  

It is reasonable to expect that this distinction
and kinematical character of the discreteness survives in the quantum 
theory. Interpretation II is consistent with usual 
quantum theory, {\em and}  with this idea.  It is therefore a sensible
hypothesis for extending quantum theory to the general-relativistic regime.

To be more specific, consider the particle on a circle, in a system where
angular momentum is not conserved: 
\begin{equation}
\frac{dp_\varphi(t)}{dt}=-\frac{\partial H_0(\varphi,p_\varphi)}{\partial \varphi}\ne 0.  
\label{pphi}
\end{equation} 
In the quantum theory, different operators are related 
to the measurement of the momentum of the particle:
the Schr\"odinger operator
\begin{equation}
\hat p_\varphi=-i\hbar \frac{d}{d \varphi} 
\label{sc}
\end{equation} 
and the one-parameter family of Heisenberg operators
\begin{equation}
(\hat P_\varphi)_t=e^{i\hbar \hat H_0t}\ \hat p_\varphi\ e^{-i\hbar \hat H_0 t}.  
\label{hs}
\end{equation} 
If we describe this system using a general-covariant formalism (see for instance
\cite{libro}) then  (\ref{sc}) can be identified with a kinematical operator in $\cal K$, while 
(\ref{hs}) can be identified with an evolving-constant operator in $\cal H$.\footnote{%
More precisely, here  ${\cal K}=L_2[R^2, d\varphi dt]$ and ${\cal H}$ is the generalized
subspace of ${\cal K}$ given by the solutions of the Schr\"odinger equation. The evolving
constant operator $(\hat P_\varphi)_t$ is well defined in  ${\cal H}$ because it commutes
with $\hat H=i\hbar\frac{\partial}{\partial t}-\hat H_0$.}

Now, a measurement of the momentum of this particle yields 
quantized outcomes. This physical fact is described by a mathematical
aspect of the theory; which one?  There are two possible answers:
(a) the momentum is quantized because the operator  $\hat p_\varphi$ has discrete
spectrum, or (b)  because the operator
$(\hat P_\varphi)_t$ has discrete spectrum. These two answers are equivalent for
\emph{this} system, because the two operators have the same spectrum. 
But they are not equivalent for a \emph{generic} 
general-covariant system, where a kinematical operator like $\hat p_\varphi$ 
and an evolving-constant like $(\hat P_\varphi)_t$  may have different spectra.  
If quantum discreteness is a 
kinematical effect that has nothing to do with the
dynamics, it 
seems to me that the correct physical answer that we have to  
take over to general-relativistic physics is the first: momentum is
quantized because the kinematical operator $\hat p_\varphi$ has
discrete spectrum. This leads to interpretation II.

(ii) {\it Gauge invariance.} The common argument presented in favor of  
interpretation I is gauge 
invariance. The argument goes as follows. Only gauge-invariant
quantities have physical meaning, and kinematical
quantities in $\cal K$ are not gauge invariant. Therefore the
spectrum of a partial observables, which is an operators on $\cal K$, 
cannot be physically meaningful. 

As clarified in  \cite{dirac}, the reason we must assume that
only the gauge-invariant  quantities can be measured and 
predicted is that the equations of motion do not determine the evolution of the 
non-gauge-invariant ones. The only quantities whose evolution is
well determined are the Dirac observables.   
This fact is taken into account within interpretation II, where 
transition amplitudes describe gauge-invariant correlations, 
and all predictions are indeed gauge-invariant. 

But in physics we utilize 
quantities that we measure but cannot predict: these are
the independent variables with respect to which we express evolution.
In non relativistic physics, the prototype of these quantities is the
time variable $t$.  In general-relativistic physics there is no preferred
time variable and \emph{any} physical quantity can play the role of independent
variable. Therefore there is nothing wrong in referring to quantities
that are not themselves Dirac observables. (This is done also within
interpretation I: an example is the parameter 
$\tau$ that parametrizes an evolving constants).  We just have to do so properly,
respecting the overall gauge-invariance of the theory and its predictions. 

There would be a contradiction if we had different spectral 
predictions in different gauges. But this is not the case, because 
the spectra of diff-related operators are the same. 
The requirement that spectra and transition 
amplitudes be gauge invariant  is therefore  implemented within 
interpretation II. Quantization of the 
partial observables is  not in contradiction 
with gauge invariance.  \\ 

(iii) {\it Gauge fixing.} In classical general relativity, the evolution of quantities that depend 
on the spacetime coordinates $x^\mu$ is under-determined by the equations
of motion.    This fact can be interpreted in two distinct but 
physically equivalent ways. (See for instance \cite{ec3}.) Accordingly, the 
spacetime coordinates $x^\mu$ 
can be given two different meanings, both consistent and viable.  

According 
to the first, the coordinates $x^\mu$ are irrelevant mathematical labels that 
can be changed at will.  The only quantities that have a physical interpretation 
are those that are independent on the choice of these coordinates.  This is the interpretation 
which is is most commonly considered in quantum gravity.   

According to the second, 
the coordinates $x^\mu$ describe physical position with respect to a physical
reference system whose dynamics we do not care describing. Then the 
under-determinacy of the evolution simply reflects the fact that we are neglecting
the evolution equations of the matter forming the physical reference system. 
In this case coordinate-dependent quantities represent  
gauge-invariant observables of a larger system where we have
\emph{gauge-fixed} the coordinates to some physical value.  In other words, the
gauge-invariant gravitational degrees of freedom on a physical
reference system are described by the same variables as the
gauge-dependent gravitational field variables.  

In simpler words, we can gauge-fix the coordinates 
by choosing them to be determined by chosen physical 
rods and clocks. Then non-diff-invariant observables in 
the pure gravity theory correspond 
precisely to diff-invariant observables in the matter+gravity theory.   
This is the analog of the fact that the Maxwell potential $A_\mu$ describes a 
physically observable quantity, if we work in a formalism in which 
we have entirely fixed the gauge. 

The fact that this is possible in the classical theory suggests 
that the same could happen in the quantum theory. That is, 
it is reasonable to expect that the gauge-invariant geometrical operators 
have the same mathematical form as the  gauge dependent ones in pure 
gravity. And therefore the same spectrum. This expectation is reinforced 
by the following consideration. \\

(iv) {\em Commutator algebra.} Discreteness depends on the 
commutation structure of the relevant geometrical quantities. 
Such structure does not change among different physical versions of 
these geometrical quantities. For instance, area elements of intersecting 
surfaces do not commute. Their algebra does not depend on how we 
have defined the surfaces. 

Compare this with the angular momentum in non relativistic quantum theory: angular
momentum is always quantized, with the same eigenvalues,
irrespectively on whether it is the angular momentum of an atom, a proton, or
a molecule, in spite of the fact that the various angular momentum 
operators are different in the different cases.  The reason is of course 
that the angular momentum functions may be different, but their $SO(3)$ 
commutator algebra is the same.   Similarly, the commutation structure of 
the components of, say, the area of any physical object must be  
dictated by the geometry of the  gravitational field, not by specific 
features of the object whose area is considered.\\

(iv) {\em Clock-dependence of the discreteness.}
Finally, let's reconsider interpretation I. Notice that in this framework the discreteness 
of the spectrum may well  depend  on the choice of the clock. One 
should therefore distinguish between two possibilities:  that discreteness of a 
quantity $f$ is lost with \emph{any} clock, or that it is lost with \emph{some} 
clock.  

The first case appears unlikely in a realistic situation, in 
the light of the discussion of the example in Sec.\,III.  If \emph{some} compact 
flow exist at all on the physical phase space, then \emph{some} complete
observable will give it and will likely have discrete spectrum.  

What is then the
interpretation of the second case?  Suppose a partial observable $f$ 
has discrete spectrum if measured at the time determined by a
clock $t_1$, but continuous spectrum if measured at the time determined 
by a clock $t_2$.  What is the physical interpretation of this situation? 
Isn't this an indication that the procedure for determining $t_2$ disturbs 
quantum mechanically the determination of $f$?  

\section{Conclusion}

The paper \cite{DT} is interesting because it points out cases
where two interpretations of general-covariant quantum theory 
can yield different results.  These cases should be
better understood and the relative merits and open questions in the two
interpretations deserve to be better investigated. 

If interpretation I is physically correct, then we do not have a hard theorem to the effect that
LQG predicts physical discreteness, but we have a number of compelling 
plausibility arguments. Counter examples appear to depend on pathologies that
do not have an obvious correspondence in the theory. 

If interpretation II is correct, then discreteness of LQG is immediate and compatible with gauge invariance.   The interpretation II appears to me not only more manageable to extract physics from the theory \cite{prop}, but also more plausible on physical ground.
\vskip.5cm
\centerline{------}
\vskip.5cm
Thanks to Bianca Dittrich, Matteo Smerlak, Simone Speziale and Thomas Thiemann for valuable comments.

\end{document}